\documentclass[aps,prd,twocolumn,superscriptaddress,showpacs,amsmath,amssymb]{revtex4}
\usepackage{graphicx,epsf}
\usepackage[]{latexsym}
\newcommand{\be}{\begin{eqnarray}}
\newcommand{\ee}{\end{eqnarray}}
\newcommand{\bea}{\begin{eqnarray}}
\newcommand{\eea}{\end{eqnarray}}

\renewcommand{\H}{\hat H}
\renewcommand{\d}{\mbox{{\rm d}}}
\def\comment#1{}

\usepackage{color}

\definecolor{darkred}{rgb}{.8,0,0}

\definecolor{darkblue}{rgb}{0,0,.7}

\definecolor{darkgreen}{rgb}{0,.7,0}

\begin{document}

%
%
%%%%%%%%%%%%%%%%%%%%%%%%%%%%%%%%%%%%%%%%%%%%%%%%%%%%%%%%%%%%%%
\title{Dissipation and quantization for composite systems}
%'t~Hooft quantization proposal for interacting systems}
%%%%%%%%%%%%%%%%%%%%%%%%%%%%%%%%%%%%%%%%%%%%%%%%%%%%%%%%%%%%%%

%
\author{Massimo~Blasone}
\email{mblasone@unisa.it}
\affiliation{INFN, Gruppo Collegato di Salerno and DMI, Universita'
di Salerno, Fisciano (SA) - 84084 Italy}
\author{Petr~Jizba}
\email{jizba@physik.fu-berlin.de}
\affiliation{ITP, Freie Universit\"{a}t Berlin, Arnimallee 14
D-14195 Berlin, Germany}
\affiliation{FNSPE, Czech Technical
University { in Prague}, B\u{r}ehov\'{a} 7, 115 19 Praha 1, Czech Republic}
\author{Fabio~Scardigli}
\email{fabio@phys.ntu.edu.tw}
\affiliation{Leung Center for Cosmology and Particle Astrophysics
(LeCosPA), Department of Physics, National Taiwan University, Taipei
106, Taiwan} \affiliation{Yukawa Institute for Theoretical Physics,
Kyoto University, Kyoto 606-8502, Japan}
\author{Giuseppe~Vitiello}
\email{vitiello@sa.infn.it}
\affiliation{INFN, Gruppo Collegato di Salerno and DMI, Universita'
di Salerno, Fisciano (SA) - 84084 Italy}
%
%
%\date{\today}
%
%
\begin{abstract}
{ In the framework  of 't~Hooft's quantization proposal, we show how
to obtain from the composite system of two classical Bateman's oscillators
a quantum isotonic
oscillator.} In a specific range of parameters, such a system can
be  interpreted as a particle in an effective magnetic field,
interacting through a spin-orbit interaction term.
{ In the limit of a large separation from the
interaction region one can} { describe the system in terms of two irreducible elementary
subsystems which correspond to two independent quantum harmonic oscillators.}
\end{abstract}

%\\
%``Nothing is more beautiful than to have\\
%a clear idea of the causes and of the effects."\\
%A.Einstein}
%
%
%
\maketitle

%%%%%%%%%%%%%%%%%%%%%%%%%%%%%%%%%%
\section{Introduction}\label{SEc1}
%%%%%%%%%%%%%%%%%%%%%%%%%%%%%%%%%%

In a series of papers~\cite{hooft1,hooft2} Gerard 't Hooft
has put forward the conjecture that the origin
of the quantum mechanical nature of our world is in the
dissipation of information which { should} occur at very high energies (Planck scale) in a
regime of deterministic dynamics.

%He has considered classical,
%deterministic, dissipative systems and has shown that by imposing
%constraints on the solutions so to introduce information loss {\em
%``an apparent quantization of the orbits which resembles the quantum
%structure seen in the real world''} \cite{hooft2} is obtained.

The basic observation of 't~Hooft is that there exists a
family of  dynamical systems that
can be described by means of Hilbert space techniques without
loosing their deterministic character. Only after enforcing certain
constraints expressing information loss, one obtains {\em bona fide}
quantum systems. In this picture the quantum states of actually
observed degrees of freedom (observables) can be identified with
equivalence classes of states that span the original (primordial)
Hilbert space of truly existing degrees of freedom (be-ables). It is
important to remark that be-ables are not referring to
conventional macroscopic variables, such as a pointer on a detection
device, but rather to a set of what 't~Hooft calls ``primordial''
variables. Conventional variables, like mass, energy, position,
etc., are viewed as emergent (non-primordial) degrees of freedom
that are described in terms of states in a quotient Hilbert space.

%'t~Hooft's work suggests that the deeper an explanation of QM in
%terms of energy scale is, the more remote from immediate experience
%are the entities (be-ables) to which it must refer.
%
%'t~Hooft was instrumental in providing explicit examples of be-able
%systems that give rise at a macroscopic level to a genuine quantum
%behavior~\cite{hooft1,hooft2}. For logical consistency of the entire
%scheme it is, however, important to demonstrate that any emergent
%composite quantum system has also its subsystems (i.e., its
%identifiable subparts) quantal. An explicit construction of such a
%model has been so far a missing element in the scheme. It is our aim
%here to fill this gap by providing an explicit model of this type.

%
%'t Hooft proposal is interesting not only from the perspective of the foundation
%of Quantum Mechanics and the debate about its interpretation (from the Copenhagen one
%to, e.g. Nelson's stochastic quantization procedure \cite{s}) but also from
%the perspective of the discussion ``classical vs. quantum'', where the Bell's inequality
%issue is commonly raised. Without mentioning possible implications in the
%most physically important issue of quantization of gravity. We will not enter in any of
%such discussions in this paper, but these issues are certainly on the background
%of the topic discussed in the present paper.

't Hooft proposal has been discussed in a number of papers~\cite{Biro,benerjee,elze,elze2,halliwell} and
explicit examples of the above scheme have been
constructed{, eg. in Refs.}~\cite{hooft2,blasone1,blasone2,scard,Jizba,JizbaII,JizbaIII}.
In particular, in Ref.~\cite{blasone1} the Hamiltonian for a couple
of classical damped-amplified oscillators~\cite{celeghini92}, also
known as Bateman's { dual oscillator} (BO), has
been shown to be a suitable deterministic system giving rise to a genuine quantum
harmonic oscillator, once the information loss condition had been implemented.

In the explicit realizations of 't Hooft quantization scheme so far considered, only
non-interacting quantum systems have been constructed. It is thus important
to study deterministic systems which would give rise to interacting quantum systems,
after appropriate constraints are imposed.
In this paper, we accomplish this task and consider the composite (deterministic) system
 made  of two couples of classical
damped-amplified oscillators. Of course, the results of
Ref.~\cite{blasone1} apply to each of the Bateman's { dual}
oscillators { separately.} We  show that the condition of
information loss { (ILC)} when applied to the global system
composed { of} two { BO} determines  the form of the
interaction of the resulting quantum system, which turns out to be a
quantum isotonic oscillator~\cite{Landau}. In a specific range of
parameters, the system can be  also interpreted as a particle in an
effective magnetic field interacting through a spin-orbital
interaction term.

%Dissipation (information loss) thus turns out to
%affect the form of the interaction among the components of a system
%and of its quantum behavior in a determinant way.

The plan of the paper is the following. As preliminaries, in the
Sections II and III we summarize the main features of 't Hooft
proposal and of the formalism introduced in Ref.~\cite{blasone1},
respectively. This will turn out to be useful for the subsequent
discussion in Section IV, where we show how the constraint of
information loss imposed on the global system dictates the interaction
between the component subsystems. Section V is devoted to final
remarks and possible avenues for future investigations.
Section VI contains the conclusions. Some further mathematical details are
confined into the Appendices.

\section{'t~Hooft's quantization scheme}\label{SEc2}
%%%%%%%%%%%%%%%%%%%%%%%%%%%%%%%%%%%%%%%%%%%%%%%%%%%%%%%%%%%%%%%%%%

In this section we briefly outline 't Hooft's continuous-time
quantization proposal~\cite{hooft2}. { To this end we} consider
{ the dynamics at the primordial, deterministic,
level as} described by the equations
\bea \dot{q}_i = f_i({\bf q})\, , \label{1.2} \eea
with a vector field $f_i$ on some configuration space $Q \subseteq
{\mathbb{R}}^n$. The system (\ref{1.2}) is generally non-Hamiltonian
but it  can be lifted~\cite{pontryagin62} to a Hamiltonian system on
the phase space $\Gamma = Q\times {\mathbb{R}}^n$, provided one
defines the Hamiltonian $H: \Gamma \mapsto {\mathbb{R}}$ as
\bea H=\sum_i p_i f_i({\bf q}) + g({\bf q})\, . \label{hc} \eea
Here $({\bf p},{\bf q}) = (p_1, \ldots, p_n, q_1, \ldots, q_n)$
denote the canonical coordinates on $\Gamma$ and $g$ is certain
function on $Q$ to be discussed shortly. Note that, due to the
particular (linear) form of the Hamiltonian in the $p_i$ variables,
the system described by Eq.(\ref{1.2}) is autonomous in the $q_i$ variables.

Besides, we have:
\begin{eqnarray}
q_i(t + \Delta t) &=& q_i(t) + f_i({\bf q})\Delta t + \frac{1}{2}
f_k({\bf q} )\frac{\partial^2 H}{\partial p_i \partial
q_k}(\Delta t)^2  \nonumber \\[1mm]
&+& \ \cdots  { \;\;  = \  F_i({\bf q}(t),\Delta t)}\, , \label{1.3}
\end{eqnarray}
where $F_i$ is some function of ${{\bf q}}(t)$ and $\Delta t$ but
not ${\bf p}$. Since (\ref{1.3}) holds for any $\Delta t$ we get the
Poisson bracket
\bea \{ q_i(t'), q_k(t)\} = 0\, , \label{1.4}  \eea
which holds for any $t$ and $t'$.

Because of the autonomous character of the dynamical equation
(\ref{1.2}) one can define a formal Hilbert space $\mathcal{H}$
spanned by the states $\{|{\bf q}\rangle; {{\bf q}} \in
{\mathbb{R}}^n\}$, and associate with $p_i$ the operator ${\hat p}_i
= -i\partial/\partial q_i$. It is not difficult to see that the
generator of time translations, the Hamiltonian operator, of the
form ${\hat H} = \sum_i {\hat p}_i f_i({\hat{\bf q}}) + g({\hat{\bf q}})$
generates the {\em deterministic} evolution equation
(\ref{1.2}) (see footnote \cite{foot1}).
Indeed, we firstly observe that because ${\hat H}$ is
generator of time translations then in the Heisenberg picture
\bea
{\hat q}_i (t + \Delta t) = e^{i\Delta t \hat{H}} {\hat q}_i (t)
e^{-i\Delta t \hat{H}}\, , \label{1.5} \eea
which for infinitesimal $\Delta t$ implies
\bea
&&{\hat q}_i (t + \Delta t) -  {\hat q}_i (t) = i \Delta
t[\hat{H}, q_i(t)]\, ,\nonumber \\[1mm]
\Rightarrow &&\dot{\hat q}_i\,  = f_i(\hat{\bf q})\,.  \label{1.6}
\eea
On the other hand, for arbitrary finite $\Delta t$ we have from
(\ref{1.5})
\bea {\hat q}_i (t + \Delta t) &=& \sum_{m=0}^{\infty} \frac{1}{m!}
[\hat{H},[\hat{H},[\cdots [\hat{H},{\hat q}_i (t)]]\cdots]]\nonumber
\\[1mm]
&=& \tilde{F}_i({\hat{\bf q}}(t),\Delta t)\, .
 \label{1.7} \eea
On the first line $\hat{H}$ appears in the generic term of the sum
$m$ times. On the second line $\tilde{F}_i$ is some function of
${\hat{\bf q}}(t)$ and $\Delta t$ but not ${\hat{\bf p}}$ which
immediately implies that
\bea [{\hat q}_i(t), {\hat q}_j(t')] = 0\, , \label{1.8} \eea
for any $t$ and $t'$ (this in turn gives $F_i = \tilde{F}_i$).
Result (\ref{1.8}) shows that the Heisenberg equation of the motion
for ${\hat q}_i(t)$ in the ${\bf q}$-representation is identical
with the $c$-number dynamical equation (\ref{1.2}). This is because
${\hat q}_i(t+ \Delta t)$ and ${\hat q}_i(t)$ commute, and hence
${\hat q}_i(t+ \Delta t)$,  ${\hat q}_i(t)$, $f_i(\hat{\bf q})$ and
also $\d {\hat{q}}_i(t)/\d t$ can be simultaneously diagonalized. In
this diagonal basis we get back the $c$-number autonomous equation
(\ref{1.2}).   From the Schr\"{o}dinger picture point of view this
means that the {\em state vector} evolves smoothly from one {\em
base vector} to another (in Schr\"{o}dinger picture base vectors are
time independent and fixed). So at each instant the state vector
coincides with some specific base vector. Because of this, there is
{\em no} non-trivial linear superposition of the state vector in
terms of base vectors and hence {\em no} interference phenomenon
shows up when measurement of ${\bf q}$-variable is performed.In
other words, the operators ${\hat q}_i$ evolve deterministically
even after ``quantization''.

Dynamical variables fulfilling Eq.(\ref{1.8}) were first considered
by Bell~\cite{Bell} who called them \emph{be-ables} as opposed to
observed dynamical variables which in QM are called
\emph{observables}.

The Hamiltonian (\ref{hc}) is unbounded from below. This fact is not
really a problem since the  dynamics of be-ables is actually
described by Eq.(\ref{1.2}). However, at  the observational
(emergent) level Hamiltonians are key objects { as they are
tightly connected with the concept of energy} of the system and {
so they} must be bounded from below.
%The concept of lower bound is in 't~Hooft proposal just
%an emergent property formed during the coarse graining of the
%be-able degrees of freedom down to the observational ones.

In order to introduce a lower bound for the Hamiltonian $\hat {H}$,
we consider a positive function of the ${\hat q}_i$ alone,
$\rho({\hat q})$, commuting with $\hat {H}$:  $[\hat{\rho}, \hat {H}
] = 0$. We can then write
\bea
&&\mbox{\hspace{-9mm}}\H \ = \ \H_+ - \H_-, \nonumber \\
&&\mbox{\hspace{-9mm}}\H_+ \ = \
\frac{1}{4\hat{\rho}}\left(\hat{\rho} + \H \right)^2 ,\quad \H_- \ =
\ \frac{1}{4\hat{\rho}}\left(\hat{\rho} - \H \right)^2 ,
\label{2.5}\eea
where $\H_+$ and $\H_-$ are positive-definite operators satisfying
\bea [\H_+,\H_-] = [\hat{\rho},\H_{\pm}] = 0 . \label{comm} \eea
One then requires  that the states at the observational level, also
called {\em physical} states $|\psi\rangle_{phys}$, must satisfy the
constraint~\cite{hooft1,hooft2}
\bea  \hat{H}_- |\psi\rangle_{phys} = 0 \, , \label{1.13} \eea
which will be henceforth referred to as the \emph{information loss
condition} (ILC). This equation identifies the states that are still distinguishable
at the observational scale.

It is interesting to interpret, in full generality, such a
constraint as a ``coarse-graining" operation induced by an operator
$\hat{\Phi}$ describing  the process of information loss occurring
when passing from the be-able to the observational level (see
Section \ref{SEc5} for further remarks on $\hat{\Phi}$).
% \footnote{ We should
%point out explicitly that the concept of ``coarse graining" stands
%also for ``loss of information", or for the more physical concept of
%``dissipation". From a macroscopic point of view (thermodynamical
%point of view), dissipation is nothing else but the passage of
%energy from some ordered degrees of freedom to some disordered
%degrees of freedom. However the concept of ``dissipation" could
%embody an idea of ``flow of time" which, at a first sight, seems to
%be missed in a ``coarse graining" procedure. Actually, this is not
%true if we think that the coarse graining procedure is applied not
%only along energy scales but also along the corresponding time
%scales. When we pass from the be-ables level $(E_P, \Delta t_P)$ to,
%say, the observational level $(E, \Delta t)$, we change the
%resolution of our time scale from $\Delta t_P$ to $\Delta t \gg
%\Delta t_P$. During the (long) time $\Delta t$ many primordial
%trajectories described by Eq.(\ref{1.2}) may have reached their
%limit cycles, and therefore the information about those primordial
%trajectories may have been lost (at the time scale $\Delta t$) in an
%irreversible way. In other words, the initial condition of a given
%trajectory cannot be retraced back, after a while. Hence this
%process, from a macroscopic point of view (observables), can be
%called, with appropriate language, ``dissipation" (or ``loss of
%information").}.
In the case of Eq.(\ref{1.13}), we have
\bea \hat{\Phi} =\hat{H}_-\,. \label{1.12} \eea
Such  a constraint is, according to Dirac's
classification~\cite{sunder}, a first class primary constraint
because $[\hat{\Phi},\hat{\Phi}] = 0$ and $[\hat{\Phi},\hat{H}] =
0$. First-class constraints generate gauge transformation and
produce equivalence classes of states~\cite{sunder}, which are
generally non-local. States belong to the same class, even if
space-like separated,  when they can be transformed into each other
by gauge transformations  generated by  $\hat{\Phi}$. Let
$\mathcal{G}$ be the group of these gauge transformations.

The above equivalence classes represent the physical states, namely
the ones accessible at observational level in Quantum Mechanics.
Denoting with $\mathcal{O}$ the space spanned by the observables, we
identify it with the quotient space
\bea
\mathcal{O}=\mathcal{H}_c/\mathcal{G}\, .
\label{2.4}
\eea

From Eq.(\ref{1.13}), we see that physical states have positive energy
spectrum, since
\bea \H|\psi\rangle_{phys} = \H_+|\psi\rangle_{phys} =
\hat{\rho}|\psi\rangle_{phys}\, .  \eea
Thus, in the Schr{\"o}dinger picture the equation of motion
\bea \frac{\d}{\d t}|\psi_t\rangle_{phys} = -i
\H_+|\psi_t\rangle_{phys}\, , \label{2.6} \eea
has only positive frequencies on physical states.
%
%Coarse-graining procedure which we have discussed so far was
%implemented in an operatorial mode of description of the be-able
%dynamics. As a result we have obtained at an observational
%scale a Schr\"{o}dinger equation with a ``well behaved" (i.e.,
%bounded from below) emergent Hamiltonian.

In Ref.~\cite{hooft2} 't~Hooft observed  that  when the dynamical
equations (\ref{1.2}) describe  configuration-space chaotic
dynamical systems, the equivalence classes could be related to their
stable orbits (e.g., limit cycles). Information loss is responsible
for clustering of trajectories to equivalence classes: after a while
one cannot retrace back the initial conditions of a given
trajectory; one can only say to which attractor the evolution leads.
Applications of the the outlined canonical scenario were given,
e.g., in Refs.~\cite{blasone1, blasone2, scard}.

{
%%%%%%%%%%%%%%%%%%%%%%%%%%%%%%%%%%%%%%%%%%%%%%%%%%%%%%%%%%%%%%%
\section{Two Bateman's dual oscillators: local ILC}\label{SEc3}
%%%%%%%%%%%%%%%%%%%%%%%%%%%%%%%%%%%%%%%%%%%%%%%%%%%%%%%%%%%%%%%
%
}

{ A couple} of damped-amplified harmonic oscillators (i.e. a Bateman's
{ dual} oscillator)
was considered in Ref.~\cite{blasone1} as an explicit example yielding 't
Hooft's be-able dynamics.
{ There, the ILC --- Eq.~(\ref{1.13}), gave rise to a genuine one-dimensional
quantum harmonic oscillator.}

In the present section we consider two { BO}, labeled
by the index $i=A,B$:
\begin{eqnarray} && m_i {\ddot{x}}_i + \gamma_i {\dot{x}}_i +
\kappa_i x_i = 0 \, ,  \\[2mm]
&& m_i{\ddot{y}}_i - \gamma_i {\dot{y}}_i + \kappa_i y_i = 0\, ,
\label{4.1} \end{eqnarray}
where $m_i = (m_A,m_B)$, $\gamma_i = (\gamma_A, \gamma_B)$ and
$\kappa_i = (\kappa_A, \kappa_B)$. The $y_i$--oscillator is the
time--reversed image of the $x_i$--oscillator. In the following, as
a preliminary to the discussion of section \ref{SEc4}, we will
closely follow the treatment presented in Ref.~\cite{blasone1},
where the reader can find details here omitted for brevity.

The  Lagrangian for the $i$th-couple of oscillators is
\bea {\mathcal{L}}_i = m_i {\dot{x}}_i {\dot{y}}_i +
\frac{\gamma_i}{2} \ (x_i {\dot{y}}_i - {\dot{x}}_i y_i) - \kappa_i
\ {x}_i {y}_i \, , \label{xylag} \eea
and the conjugated momenta are
\bea
p_{x_i} &=& \frac{\partial {\mathcal{L}}_i}{\partial {\dot{x}}_i} =
m_i {\dot{y}}_i  -  \frac{1}{2} \ \gamma_i  y_i\, , \nonumber \\
p_{y_i} &=& \frac{\partial {\mathcal{L}}_i}{\partial {\dot{y}}_i} =
m_i {\dot{x}}_i  +  \frac{1}{2} \ \gamma_i  x_i \, .
\eea
Therefore the Hamiltonian for $i$th oscillator reads
\bea
H_i =  \frac{1}{m_i}p_{x_i}p_{y_i} &+& \frac{\gamma_i}{2 m_i} (y_i p_{y_i} - x_i p_{x_i}) \nonumber \\
&+& \left(\kappa_i - \frac{\gamma_i^2}{4 m_i}\right)x_i y_i \, .
\label{ham1}
\eea
In order to show that $H_i$ belongs to the class of 't~Hooft's Hamiltonians,
it is convenient to reformulate the former system in a rotated coordinate
frame, i.e.
\begin{eqnarray*}
x_{1i} = \frac{x_i+ y_i}{\sqrt{2}}\, , \,\,\,\,\ x_{2i} =
\frac{x_i-y_i}{\sqrt{2}}\, . \label{4.1a}
\end{eqnarray*}
In these new coordinates the $i$th Lagrangian has the form
\begin{eqnarray}
{\mathcal{L}}_i = \frac{m_i}{2}  ({\dot{x}}_{1i}^2 - {\dot{x}}_{2i}^2) &+&
\frac{\gamma_i}{2} ({\dot{x}}_{1i} {x}_{2i} - {x}_{1i} {\dot{x}}_{2i})  \nonumber \\
&-& \frac{\kappa_i}{2} ({x}_{1i}^2 - {x}_{2i}^2)  \, .
\end{eqnarray}
For the new canonical momenta $p_{\alpha i} = \partial {\mathcal{L}}/ \partial {\dot{x}_{\alpha i}}$,
where $x_{\alpha i} \equiv (x_{1i},x_{2i})$ and $p_{\alpha i} \equiv (p_{1i},p_{2i})$, we obtain
\begin{eqnarray}
p_{1i} &=& m_i {\dot{x}}_{1i}  +  \frac{1}{2} \ \gamma_i  x_{2i}\, ,  \\
p_{2i} &=& - m_i {\dot{x}}_{2i}  -  \frac{1}{2} \ \gamma_i  x_{1i}\, ,
\end{eqnarray}
and thus the corresponding $i$th Hamiltonian reads
\begin{eqnarray}
H_i &=& \frac{1}{2m_i} ({p}_{1i}^2 - {p}_{2i}^2) - \frac{\gamma_i}{2m_i}
({p}_{1i} {x}_{2i} + {p}_{2i} {x}_{1i}) \nonumber \\
&+& \frac{1}{2} \left(\kappa_i - \frac{\gamma^2_i}{4m_i} \right) ({x}_{1i}^2 - {x}_{2i}^2) \,  .
\label{4.1c}
\end{eqnarray}
The algebraic structure for the total system $H_T = H_A + H_B$
is the one of $su(1,1)\otimes su(1,1)$. Indeed,
from the dynamical variables ${{p}}_{\alpha i}$ and ${x}_{\alpha i}$
one may construct the functions
\begin{eqnarray}
J_{1 i} &=& \frac{1}{2m_i \Omega_i} \ p_{1i} p_{2i} -  \frac{m_i \Omega_i}{2}
\ x_{1i} x_{2i}\, , \nonumber
\\
J_{2i} &=& \frac{1}{2} ({p}_{1i} {x}_{2i} + {p}_{2i} {x}_{1i}) \, , \nonumber
\\
J_{3i} &=& \frac{1}{4m_i \Omega_i} \left(p_{1i}^2 + p_{2i}^2 \right)
+ \frac{m_i\Omega_i}{4}\ \left(x_{1i}^2  + x_{2i}^2 \right)\, ,
\end{eqnarray}
where $\Omega_i = \sqrt{\frac{1}{m_i}
(\kappa_i-\frac{\gamma_i^2}{4m_i})}$, and $\kappa_i
>\frac{\gamma_i^2}{4m_i} $. Applying now the canonical Poisson brackets $\{x_{\alpha i}, p_{\beta j}\}
= \delta_{\alpha \beta}\delta_{ij}$ we
obtain the Poisson's subalgebra
\begin{eqnarray}
&&\{ J_{2i},J_{3i}\} = J_{1i}\, , \,\,\,\, \{ J_{3i}, J_{1i} \} = J_{2i}\, ,\nonumber \\[2mm]
&&\{ J_{1i},J_{2i} \} = - J_{3i}\, , \,\,\,\, \{J_{\alpha i}, J_{\beta j} \}|_{i\neq j}  = 0 \, .
\label{poissons}
\end{eqnarray}
The algebraic structure (\ref{poissons}) corresponds to
$su(1,1)\otimes su(1,1)$ algebra. The quadratic Casimirs for the
algebra (\ref{poissons}) are defined as
\begin{eqnarray} {\mathcal{C}}^2_i =   J_{3i}^2 - J_{2i}^2 - J_{1i}^2 \, .
\label{3.9}
\end{eqnarray}
The  ${\mathcal{C}}_i$ explicitly  read
\bea
{\mathcal{C}}_i = \frac{1}{4 m_i \Omega_i}[({p}_{1i}^2 - {p}_{2i}^2)
+ m_i^2 \Omega_i^2({x}_{1i}^2 - {x}_{2i}^2)]\, .
\eea
In terms of $J_{2i}$ and  ${\mathcal{C}}_i$ the Hamiltonians $H_i$
are given by
\begin{eqnarray}
H_i \ = \ 2  \left(\Omega_i {\mathcal{C}}_i - \Gamma_i J_{2i}   \right)\, ,
\label{4.6a}
\end{eqnarray}
where $\Gamma_i = \gamma_i/ 2 m_i$. Eq.(\ref{4.6a}) shows that $H_i$
are of the 't~Hooft form, with the ${\mathcal{C}}_i$ and $J_{2i}$
playing the role of $p$'s, and the $\Omega_i$ and $\Gamma_i$ the one
of $f({\bf q})$'s.

{ A further simplification can be achieved by introducing} the hyperbolic coordinates:
\begin{eqnarray}
&&x_{1i} \ =\  r_i \cosh u_i \, ,  \nonumber \\[2mm]
&&x_{2i} \ = \ r_i \sinh u_i \, ,  \,\,\,\,\,  r_i \in {\mathbb{R}}\, ;
u_i \in {\mathbb{R}}\, .
\label{43}
\end{eqnarray}
{ The corresponding  conjugated momenta then are}~\cite{BlasVit}
\bea
p_{r_i} &=& p_{1i}\cosh u_i + p_{2i}\sinh u_i  \, , \nonumber \\
p_{u_i} &=& p_{1i}r_i\sinh u_i + p_{2i}r_i\cosh u_i\, . \label{hypm}
\eea
In these coordinates $J_{2i}$ and ${\mathcal{C}}_i$ have a
particularly simple structure~\cite{blasone1}, namely
\begin{eqnarray}
&&{\mathcal{C}}_i  \ = \ \frac{1}{4 \Omega_i m_i}\left[ p_{r_i}^2 -
\frac{1}{r^2_i}p_{u_i}^2 + m_i^2\Omega_i^2 r_i^2\right]\, , \nonumber \\[2mm]
&&J_{2i} \ = \ \frac{1}{2}\, p_{u_i}\, . \label{44}
\end{eqnarray}

In Ref.~\cite{blasone1} it is shown that the
following canonical transformations hold:
\begin{eqnarray}
&&q_{1i} = \int \frac{dz_i \;\; m_i\Omega_i}{(4 J_{2i}^{2} + 4m_i\Omega_i {{\mathcal{C}}_i} z_i
- m_i^{2}\Omega^{2}_i z^{2}_i)^{1/2}}\, ,\nonumber \\[2mm]
&&q_{2i} = 2u_i + \int \frac{dz_i}{z_i} \, \frac{2J_{2i}}{(4
J_{2i}^{2} + 4m_i\Omega_i {{\mathcal{C}}_i}
z_i - m_i^{2}\Omega_i^{2} z_i^{2})^{1/2}}\, ,\nonumber \\[2mm]
&&p_{1i} = {\cal C}_i\, , \;\;\;\;\ p_{2i} = J_{2i} \, , \label{can1}
\end{eqnarray}
with $z_i=r^2_i$. One has $\{{\bf q}_{i},{\bf p}_i\} =1$, and the
other Poisson brackets vanishing.

%Secondly, the role of $J_{2}$ is
%very exceptional in the case of Bateman's system. As widely argued in
%Refs.~\cite{blasone1,celeghini92}, the generator $J_{2}$, which controls the dissipative part
%of the dynamics, behaves like an entropy and hence can be
%related to an information measure (entropy) of the system (\ref{4.6a}). Thus, it
%serves as a natural candidate for 't~Hooft's ILC. We shall return to
%this point shortly.

Let us now consider the {\em operatorial}  description of this
be-able dynamics. To this end we promote all the relevant
quantities, $H_i$, ${\cal C}_i$, $J_{2i}$ and ${\bf q}_i$ to
operators.
%We can then properly speak about their commutators, and
{ Note that $({\bf q}_i)$  and separately $({\cal C}_i, J_{2i})$ are two
independent sets of be-ables.}
%Moreover, for spectral purposes it is often
%convenient~\cite{celeghini92,Perelomov} to consider instead of the
%(\ref{3.9}) the Casimir
The Casimir operators are~\cite{celeghini92,Perelomov}:
\begin{eqnarray} \hat{\mathcal{C}}^2_i =  \hat{J}_{3i}^2 -\hat{J}_{2i}^2 -
\hat{J}_{1i}^2 + \mbox{$\frac{1}{4}$} \, ,
\label{3.10}
\end{eqnarray}
{ where the factor $1/4$ was introduce for convenience.}
Following 't~Hooft, we now write the above Hamiltonians in the form
\begin{eqnarray}
&&\H_i = \H_{i+} - \H_{i-}\, , \nonumber \\[1mm]
&&\H_{i+} = \frac{1}{4\hat{\rho_i}}(\hat{\rho_i} + \H_i)^2 , \quad
\H_{i-} = \frac{1}{4\hat{\rho_i}}(\hat{\rho_i} - \H_i)^2\, .
\end{eqnarray}
Choosing $\hat{\rho_i}=2\Omega_i \hat{\cal C}_i$, and taking
$\hat{\cal C}_i>0$ (this can be done, because $\hat{\cal C}_i$ are
constants of motion), the splitting reads
\bea \H_{i+}&=&\frac{(\H_i + 2\Omega_i \hat{\cal C}_i)^2}{8\Omega_i
\hat{\cal C}_i} = \frac{1}{2\Omega_i \hat{\cal C}_i}(2\Omega_i
\hat{\cal C}_i - \Gamma_i \hat{J}_{2i})^2\, ,
\nonumber \\
\H_{i-}&=&\frac{(\H_i - 2\Omega_i \hat{\cal C}_i)^2}{8\Omega_i
\hat{\cal C}_i} = \frac{1}{2\Omega_i \hat{\cal C}_i} \Gamma_i^2
\hat{J}_{2i}^2\, . \label{4.5} \eea
Quantization emerges after the information loss condition is imposed { locally, i.e.
separately on each of the Bateman oscillators}:
\begin{eqnarray}
\hat{J}_{2i}|\psi\rangle_{phys} = 0\, , \label{III.39}
\end{eqnarray}
which defines/selects the physical states and is equivalent to
\bea
\H_{i-}|\psi\rangle_{phys}=0 \label{con} , \;\;\;\; i = A,B\, .
\eea
%
% { The same constraint
% allows us to get the lower bound for the Hamiltonian by projecting
% out the states responsible for the negative part of the spectrum.}
% %
This implies
\bea
\H_i|\psi\rangle_{phys} &=& (\H_{i+} - \H_{i-})|\psi\rangle_{phys}\, ,\nonumber\\
&=&\H_{i+}|\psi\rangle_{phys} = 2\Omega_i \hat{\cal C}_i
|\psi\rangle_{phys}\, , \eea
and
\bea 2\Omega_i \hat{\cal C}_i |\psi\rangle_{phys}\! &=&\!
\left[\frac{1}{2m_i}\left(\hat{p}_{r_i}^2+m_i^2\Omega_i^2\hat{r}_i^2\right)-
\frac{2\hat{J}_{2i}^2}{m_i \hat{r}_i^2}\right]|\psi\rangle_{phys},\nonumber \\
&=&\!\left(\frac{\hat{p}_{r_i}^2}{2m_i}+\frac{m_i}{2}\Omega_i^2\hat{r}_i^2\right)|\psi\rangle_{phys}\, .
\label{ho} \eea
Eq.~(\ref{ho}) reproduces, for  each one of the systems $A$ and $B$ { separately}, the result of
Ref.~\cite{blasone1}, namely each one of the Hamiltonians $H_A$ and $H_B$ reduces
independently to the Hamiltonian of a QM oscillator.

In Appendix 1 we discuss some properties of the physical states $|\psi\rangle_{phys}$.

\vspace{5mm}

{
%%%%%%%%%%%%%%%%%%%%%%%%%%%%%%%%%%%%%%%%%%%%%%%%%%%%%%%%%%%%%%%
\section{Two Bateman's dual oscillators: global ILC}\label{SEc4}
%%%%%%%%%%%%%%%%%%%%%%%%%%%%%%%%%%%%%%%%%%%%%%%%%%%%%%%%%%%%%%%
%
}
%
% %%%%%%%%%%%%%%%%%%%%%%%%%%%%%%%%%%%%%%%%%%
% \section{The composite system}~\label{SEc4}
% %%%%%%%%%%%%%%%%%%%%%%%%%%%%%%%%%%%%%%%%%%

{ We now enforce the ILC globally, i.e., the
condition (\ref{1.13}) will be applied on the
composite system described by the Hamiltonian $H_T = H_A + H_B$.
We will see that the global enforcement of the ILC dictates the form
of the interaction between the component subsystems.}

{ We start by writing the total  Hamiltonian as}
\bea
\mbox{\hspace{-4mm}}H_T\! &=& \!H_A + H_B\, , \nonumber \\
     &=& \!2(\Omega_A {\cal C}_A + \Omega_B {\cal C}_B) - 2(\Gamma_A J_{2A} + \Gamma_B
     J_{2B})\, .
\eea
The fact that the Casimirs ${\cal C}_i$ are constants of motion,
guaranties that, once they are chosen to be positive (as we do from
now on), they remain such at all times. Also $J_{2i}$ are constants
of motion ($\{H_i, J_{2i}\}=0$). We can therefore define new
integrals of motion:
\bea {\cal C} \ \equiv \ \frac{\Omega_A {\cal C}_A + \Omega_B {\cal
C}_B}{\Omega}\, , \quad J\ \equiv\ \frac{\Gamma_A J_{2A} + \Gamma_B
J_{2B}}{\Gamma}\, , \eea
where $\Omega$ and $\Gamma$ are quantities to be defined shortly.
Using the fact that $\Omega_i > 0$ and assuming that $\Omega > 0$,
we conclude that
\bea {\cal C}_A,\,\,{\cal C}_B>0 \quad\quad \Rightarrow \quad\quad {\cal C}>0\, . \eea
The positivity of ${\cal C}$ is guaranteed by our choice ${\cal C}_i>0$.
The Hamiltonian for the total system  is
\bea H_T=2\Omega {\cal C} - 2\Gamma J\, ,\label{HT} \eea
and it reproduces exactly the Hamiltonian of each one of the two
subsystems (cf. Eq.(\ref{4.6a})).  With the choice $\rho=2\Omega
{\cal C}$, $H_T$ can be split  as (cf. Eq. (\ref{4.5}))
\bea H_{+}&=&\frac{(H_T + 2\Omega {\cal C})^2}{8\Omega {\cal C}} =
\frac{1}{2\Omega {\cal C}}(2\Omega {\cal C} - \Gamma J)^2\, ,
\label{IV46a} \\
H_{-}&=&\frac{(H_T - 2\Omega {\cal C})^2}{8\Omega {\cal C}} =
\frac{1}{2\Omega {\cal C}} \Gamma^2 J^2\, . \label{IV46b}\eea
Note that $\rho \equiv 2\Omega {\cal C}$ is a positive integral of
motion.  ${\cal C}$ { and} $J$ are again be-ables because they are
functions of be-ables.

Now we switch to the operatorial description. Following 't~Hooft, we
impose the { ILC} on the observational scale in the form
\bea \H_-|\psi\rangle_{phys}= \hat{J}|\psi\rangle_{phys} = 0\, . \label{46}\eea
This implies
%~\footnote{Now, since ${\cal C}>0$ automatically, we proceed to
%implement the constraint $\H_-|\psi\rangle_{phys}=0$, that is
%$\hat{J}^2|\psi\rangle_{phys} =0$. From this point of view, we
%introduce only $\hat{J}=0$, and therefore we are only sure that the
%global system $\H_T$ is a quantum system (after the implementation
%of the constraint). Clearly, from the present point of view, we are
%no more guaranteed that the subsystems are quantum systems (we don't
%know if $\H_1$, $\H_2$ are bounded from below, even after the
%enforcing of the constraint). Actually, after the implementation of
%the constraint $\hat{J}=0$ we are no more able, in general, to
%identify subsystems ($\H_1$, $\H_2$) into the global hamiltonian
%$\H_T$ (see Eq. (\ref{66})). During the interaction the global system is no
%more further decomposable in quantum subsystems (a property, this, shared by
%the majority of the interacting systems). When the two subsystems are far from the interaction region
%(i.e. for large $r_A$, $r_B$), they become clearly distinguishable (see Eq. (\ref{74})).}
%
%
\bea \H_T\approx \H_ + \approx  2\Omega \hat{{\cal C}} \, , \eea
where $\approx$ indicates that operators are equal only on the
physical states. Since $\hat{J}=(\Gamma_A \hat{J}_{2A} + \Gamma_B
\hat{J}_{2B})/\Gamma$, the condition $\hat{J} \approx 0$ { indicates that there must exist
a relation between} $\hat{J}_{2A}$ and $\hat{J}_{2B}$,
which in turn implies a relation between $\H_A$ and $\H_B$. { In other words,
the global ILC establishes an interaction between the two Bateman's oscillators.} We shall
now study what kind of interaction this will induce.
Solving with respect to $\hat{J}_{2B}$,  equation (\ref{46}) gives
\begin{eqnarray}
\hat{J}_{2B} \approx -\frac{\Gamma_A}{\Gamma_B}\hat{J}_{2A}\, ,
\label{J} \end{eqnarray}
which, when substituted into $H_T$
%%
%\bea \H_T &=& \H_A + \H_B \nonumber \\ &=& 2(\Omega_A \hat{\cal C}_A
%+ \Omega_B\hat{\cal C} _B) - 2(\Gamma_A \hat{J}_{2A} +
%\Gamma_B \hat{J}_{2B}) \nonumber \\
%&=&2\Omega \hat{\cal C} - 2\Gamma \hat{J}\, , \eea
%%
yields (see Eq.~(\ref{44}))
\bea
\mbox{\hspace{-11mm}}\H_T &\approx& \H_+\nonumber \\
&\approx&
\left(\frac{\hat{p}_{r_A}^2}{2m_A}-\frac{2\hat{J}_{2A}^2}{m_A\,\hat{r}_A^2}+
\frac{1}{2}m_A\,\Omega_A^2\hat{r}_A^2\right)\nonumber \\
&+&
\left(\frac{\hat{p}_{r_B}^2}{2m_B}+\frac{1}{2}m_B\,\Omega_B^2\hat{r}_B^2\right)
- \frac{2}{m_B}\,\frac{\Gamma_A^2}{\Gamma_B^2}\,\frac{1}{\hat{r}_B^2}\,
 {\hat J}_{2A}^2\, . \label{66} \eea
Note that the emergent Hamiltonian $H_T$ in Eq.~(\ref{66})
is, by construction, bounded from below. The term inside the first parenthesis is
$2\,\Omega_A \hat{\cal C}_A$. Such a term is constant, because
$\hat{\cal C}_A$ is an integral of motion.
The second term represents a QM oscillator, while the third corresponds to a centripetal
barrier. The inverse square potential $1/r^2$ is analogous to the
centrifugal contribution in polar coordinates and one may thus
expect an exact solvability. The only difference here is that $r\in
{\mathbb{R}}$ and not merely ${\mathbb{R}}^+$. The system with the
Hamiltonian
\begin{eqnarray} H = \frac{N^2}{2}\,p_{r_B}^2 + \frac{Q^2}{2}\, r_B^2 +
\frac{R^2- N^2/4}{2r_B^2}\, , \label{58b} \end{eqnarray}
and
\begin{eqnarray*}
 r_B\in {\mathbb{R}},\;\; N,Q,R\in
{\mathbb{R}}^+ \, ,
\end{eqnarray*}
is known in quantum optics and in theory of coherent
states~\cite{Wang00,Saad04} as the {\em isotonic}
oscillator~\cite{Landau}. Its spectrum can be exactly solved by
purely algebraic means since the Hamiltonian admits a
shape-invariant factorization~\cite{Dongpei}. The energy eigenvalues
read~\cite{Dongpei,Grosche}
\begin{eqnarray}
\mbox{\hspace{-2mm}}E_{n,\mp} = {QN}\left(2n \mp {\frac{R}{N}}
+1\right)\!, \;\; n \in \mathbb{N} \, .
\end{eqnarray}
If $R/N \leq 1/2$, the potential is attractive in the origin, and
both the negative and positive sign must be taken into
account~\cite{Grosche}.  The positive sign in front of $R/N$ has to
be taken when $R/N > 1/2$. The inverse square potential is then
repulsive at the origin, so the motion takes place only in the
domain $r_B >0$.

Since in our case %(see Eq.(\ref{III42aa}) in Appendix 1)
\begin{eqnarray*}
N^2=\frac{1}{m_B}, \quad  Q^2=m_B\Omega_B^2, \quad \frac{R^2}{N^2}= \frac{1}{4}
-  \left(\frac{2\Gamma_A}{\Gamma_B}\ \!\mu_A\right)^{\!2}\! ,
\end{eqnarray*}
the actual spectrum of (\ref{66}) is
\begin{eqnarray}
&&{}\hspace{-.5cm}
E_{n,\mp}(c,\mu_A) \!=\! \Omega_B \left(2n \ \mp \ \sqrt{\frac{1}{4} \! -
\!\left(\frac{2\Gamma_A}{\Gamma_B}\ \!\mu_A\right)^{\!2}} \! + \!
1\right)\! +\! c\, ,\nonumber \\[2mm]
&&n \in \mathbb{N}\, , \label{4.7}
\end{eqnarray}
where $c$ is a  constant term due to the presence of $2\Omega_A
\hat{\cal C}_A$ in Eq.(\ref{66}) and $\mu_A \approx  \hat{J}_{2A}$.
Note that when $\Gamma_A$ is small (in particular
$2\Gamma_A\mu_A/\Gamma_B \ll 1/2$), the inverse square potential in
(\ref{66}) can be neglected and the system reduces to that of a QM
linear oscillator  with a shift term $c$. This follows also directly
from the spectrum (\ref{4.7}) provided we set $\Gamma_A = 0$ and
consider both signs. In Appendix 2 we consider the case $m_A = m_B =
M$, $\Omega_A = \Omega_B = \Omega$. Then our system reproduces the
Smorodinsky--Winternitz system~\cite{Grosche} and is related to the
Calogero--Moser system~\cite{Calogero}.

It is interesting to consider the case where  $\mu_A$ is an
imaginary number, i.e., when $\hat{J}_{2A}$ belongs to a non-unitary
realization of the $D_j$ series (see Appendix 1). Then $R/N > 1/2$,
the potential in (\ref{58b}) is repulsive, and the motion takes
place only in the domain $r_B \geq 0$. This allows to us to view
$r_B$ as a radial coordinate and the inverse square potential in
(\ref{66}) and (\ref{58b}) as a rotationally invariant interaction of
the spin-orbit type. To see this we rewrite the interaction
potential in (\ref{66}) in the form
\begin{eqnarray}
&&\mbox{\hspace{-7mm}}\H_{int} \approx
\frac{2}{m_B}\,\frac{\Gamma_A}{\Gamma_B}\,\frac{1}{\hat{r}_B}\frac{\partial
V}{\partial r_B}({\bf {\hat J}}_B \cdot {\bf {\hat J}}_A) \, , \nonumber \\[3mm]
&&\mbox{\hspace{-7mm}} V \ = \ \log r_B\, . \label{53}
\end{eqnarray}
Here we have used (\ref{J}). We now formally identify ${\bf J}_B =
i{\bf L}$ and  ${\bf J}_A = i{\bf S}$, where ${\bf L}$ plays the
role of the orbital angular momentum of the  ``particle'' $B$ and
${\bf S}$ plays the role of its ``spin''.
The interaction energy then reads
\begin{eqnarray}
H_{int} &=&
- \frac{2}{m_B}\,\frac{\Gamma_A}{\Gamma_B}\,\frac{1}{r_B}\frac{\partial
V}{\partial r_B}({\bf L} \cdot {\bf S})\nonumber \\
&\equiv&  - \frac{g}{2m_Bc^2} \,\frac{1}{r_B}\frac{\partial
V}{\partial r_B}({\bf L} \cdot {\bf S})\, ,  \label{hint}
\end{eqnarray}
where we have identified $\Gamma_A/\Gamma_B$ with $g/4 c^2$, i.e.
with a quarter of the gyromagnetic factor (on the second line in
(\ref{hint}) we have included Thomas's factor $1/2$). The function $V$
thus plays the role of the planar radial scalar electromagnetic potential.
Note that $V$ is the harmonic function in the plane, i.e., $\nabla^2
V(r_B) = \delta(r_B)$.

Eqs.(\ref{53}) and (\ref{hint}) can be written as
\bea H_{int}=-\mu
\cdot {\bf B} \eea with \bea {\bf
B}=\frac{2}{m_B}\,\frac{1}{\Gamma_B}\,\frac{1}{r_B}\frac{\partial
V}{\partial r_B}\,{\bf J}_B\, , \eea \bea \mu = -\Gamma_A{\bf J}_A =
- \Gamma_A i\,{\bf S}\, , \eea
thereby suggesting the following
interpretation of the emergent global quantum system: In the
interaction region (small $r_A$, $r_B$) ${\bf B}$ plays the r{\^o}le
of a magnetic field acting on the planar system of magnetic moment
${\bf \mu}$.  ${\bf J}_B$ is the orbital angular momentum of the
particle ``B" and it is orthogonal to the plane where the system
lies, as it is the magnetic field ${\bf B}$. Existence of the
oscillator ``A" is reflected in the constant term contribution
$2\Omega_A \hat{\cal C}_A$ and in the spin of the particle ``B'' in
the spin-orbit interaction term.

We note that also the spin ${\bf J}_A \equiv i {\bf S}$ is
orthogonal to the configuration plane. { In conclusion,} the emergent
global QM system can be interpreted as a particle in a magnetic
field, with a spin-orbit interaction term.

Far from the interaction region (large $r_A$, $r_B$), the
interaction is switched off and the asymptotic emergent QM Hamiltonian reads
\bea
&&\mbox{\hspace{-8mm}}\H_T \approx \H_+  \nonumber \\
&&\mbox{\hspace{-8mm}}\approx
\left(\frac{{\hat p}_{r_A}^2}{2m_A}+\frac{1}{2}m_A\Omega_A^2\hat{r}_A^2\right) +
\left(\frac{{\hat p}_{r_B}^2}{2m_B}+\frac{1}{2}m_B\Omega_B^2\hat{r}_B^2\right)
\label{74}
\eea
The system is still a genuine QM system because both the two
non interacting quantum harmonic oscillators
are bounded from below.
For small $r_A^2$, $r_B^2$ the
interaction term is relevant and the global system is no further
decomposable into { two independent}  subsystems.

Summarizing,  before imposing the information loss condition on the
global system, we have two non interacting, independent oscillators.
The enforcement of the dissipation constraint on the global system
gives rise to the interaction $J_{2A} \leftrightarrow J_{2B}$. So
the constraint dictates the form of the interaction. This depends on
the dissipations constants $\Gamma_A, \Gamma_B$ and can be switched
off by setting $\Gamma_A \to 0$. Dissipation thus plays a key
r\^{o}le in the interaction.

\vspace{.5cm}

%%%%%%%%%%%%%%%%%%%%%%%%%%%%%%%%%%%%%%%%%%%%%%%%%%%%%%%%%%%%%
\section{Further Remarks}~\label{SEc5}
%%%%%%%%%%%%%%%%%%%%%%%%%%%%%%%%%%%%%%%%%%%%%%%%%%%%%%%%%%%%%
%
%
{ One can draw an interesting parallel with the conclusions
reached in Ref.~\cite{MS}. There, the authors consider a deformed special relativity (DSR) model
which requires a Planck constant $\hbar$ dependent on the
energy scale. In particular, in their model $\hbar(E) \to 0$ for $E
\to E_{P}$, where $E$ { is the energy scale of the particle to which the deformed
Lorenz boost is to be applied},
%corresponds to the actual energy scale
while $E_P$ is the
Planck scale energy. The Planck scale plays in Ref.~\cite{MS} an analogous r\^{o}le as the be-able
scale in the 't~Hooft case.
The basic commutator in~\cite{MS} is written as
\be
\left[{\hat{q}}^i, {\hat{p}}^{kin}_j\right]
=\delta^i_j\left(1-\frac{E}{E_P}\right)\hbar \, ,
\label{defcomm}
\ee
{ where ${\hat{p}}^{kin}_j$ indicates the kinetic momentum operator
(i.e., ${\hat{p}}^{kin}_j  \propto \d
{\hat{q}}^j/\d t $). In this connection note that the ${\hat{p}}_i$ introduced in
Section~\ref{SEc2} is just an auxiliary variable fulfilling $[{\hat{q}}^i, {\hat{p}}_j] = \delta^i_j$
at be-able scale, while the  ${\hat{p}}^{kin}_j$  fullfils at be-able scale $[{\hat{q}}^i, {\hat{p}}^{kin}_j] =
0$ (cf. Eq.~(8)).  According to (\ref{defcomm}) the effective Planck constant runs with energy as
$\hbar(E)=\hbar (1-E/E_P)$. }
%This reminds us of the algebraic deformation thought to be the source of quantization,
% at a purely algebraic level \cite{Lichnerowicz}.
For energies $E \ll E_P$ the usual Heisenberg commutator
is recovered, but when $E=E_P$ one has $\hbar(E_P)=0$.
Hence in this model the world is classical at the Planck scale, exactly as
evoked in the 't~Hooft proposal, { provided one identifies the be-able
scale with the Planck scale.}

To be more precise, from the
viewpoint of the present paper, the operators ${\hat{q}}$,
${\hat{p}}^{kin}$ should also depend on the energy scale. In fact,
at the Planck scale they cannot represent anymore macroscopic
concepts as position or momentum, but they will likely represent
be-able degrees of freedom of unknown nature. One should thus more
correctly write
${\hat{q} }^i\left({E}/{E_P}\right)$ and
${\hat{p}}^{kin}_j\left({E}/{E_P}\right)$
where, in general, ${\hat{q}}(0)={\hat{q}}$, but
${\hat{q}}({E}/{E_P})\neq{\hat{q}}$ for $ 0 \ll E \leq E_P$ (and analog
relations hold for ${\hat{p}}^{kin}$). Surely further investigations
on the connections between DSR models and 't Hooft proposal deserve
deeper attention.

A further interesting remark is that the { 't Hooft's ILC
(\ref{1.13}) and (\ref{1.12})} accounts for a
huge information loss that happens in the transition from the
be-able scale to the observational one. Along the line of the previous observation,
we can go
beyond 't~Hooft's constraint by assuming that the ``coarse graining"
condition scales with the
energy ({ distinguishability of primordial states degrades} with the lowering of
energy scale). { A simple  model} that exhibits energy
dependence (energy-scale running) for $\hat{\Phi}$ is \cite{foot2}
\be
\hat{\Phi}_E = { \left(1- e^{-(E_P - \H_+)\H_+^{-1}}\right)}\,\hat{H}_-\, ,
\label{1.12a}
\ee
where $E$ refers to the observer's energy scale, while $E_P$ is the be-able energy scale,
which we take to be the Planck energy.
An interesting connection with the basic deformed commutator
(\ref{defcomm}) can be established by noting that, from
Eq.(\ref{defcomm}) follows that on physical states,
\be
\left(1-\frac{\H_+}{E_P}\right) =
\frac{1}{n\hbar}\left[{\hat{q}}^i,{\hat{p}}^{kin}_i\right]
\ee
and therefore $\hat{\Phi}_E$ can be rewritten as
\be
\hat{\Phi}_E =
\left(1- e^{-\,[{\hat{q}}^i, {\hat{p}}^{kin}_i] \H_+^{-1} E_P/(n\hbar)}\right)\,\hat{H}_-\, .
\ee
Here the r\^{o}le of the commutator
(\ref{defcomm}) in defining the coarse-graining operator
$\hat{\Phi}_E$ is evident. The operator $\hat{\Phi}_E$
is then, as usual, implemented as a constraint on the be-able Hilbert space
$\mathcal{H}$. Therefore at the observational \emph{energy} scale
$E$, the observed {\em physical} states { $|\psi_E\rangle_{phys}$} must
satisfy the condition
\be \hat{\Phi}_E { |\psi_E\rangle_{phys}} = 0 \, .
\label{1.13a}
\ee
This equation identifies the { be-able states} that are still distinguishable
at the observational scale $E$.
The constraint (\ref{1.13a}) is still a first class primary
constraint because $[\hat{\Phi}_E,\hat{\Phi}_E] = 0$ and
$[\hat{\Phi}_E,\hat{H}]= 0$. So again it generates gauge
transformation~\cite{sunder} and produces equivalence classes of
states which are { again generally non-local.}

Now the group of the gauge transformations generated by
${\hat{\Phi}}_E$ is a one parameter group $\mathcal{G}_E$. The
equivalence classes obtained by such gauge transformations represent
at each fixed scale $E$ the physical states (i.e., observables). So
the space of the observables will be now denoted by $\mathcal{O}_E$,
and identified with the quotient space
\be
\mathcal{O}_E=\mathcal{H}_c/\mathcal{G}_E\, .
\label{2.4a}
\ee
The quotient space $\mathcal{O}_E$ (its structure and
dimensionality) depends on the energy scale $E$. In particular, at
the level of be-ables where $E = E_P$ the constraint $\hat{\Phi}_E$
is identically zero and the space of observables is directly the
Hilbert space $\mathcal{H}$. On the other hand, when $E \ll E_P$,
e.g., at scales available to a human observer, we have $\hat{\Phi} =
\hat{H}_-$. The latter is the constraint originally considered by
't~Hooft~\cite{hooft1,hooft2}. In Section~\ref{SEc2} we have dealt
precisely with this case.}

\section{Conclusions}~\label{SEc6}
%%%%%%%%%%%%%%%%%%%%%%%%%%%%%%%%%%%%%%%%%%%%%%%%%%%%%%%%%%%%%
%
%
{
In this paper we have considered the problem of quantization of a
composite system, in the framework of the quantization scheme
proposed by G.~'t Hooft~\cite{hooft1}. { The presented analysis extends the results developed in
a previous paper~\cite{blasone1}, where only
a single Bateman's  dual oscillator was considered. In this latter case a quantum harmonic oscillator was shown to emerge
after the ILC was enforced~\cite{blasone1}}.

In the present paper, we have considered two Bateman's { dual} oscillators and
shown that in this case two possibilities arise: { one is to impose
the ILC locally, i.e., on each BO
separately, thus arriving at two independent quantum harmonic
oscillators. Another possibility is to apply the ILC globally, i.e., to the
composite system of two BO, and this in turn leads to an
interacting quantum system.} { We have worked out both possibilities
and have shown that the second option leads to an emergent quantum systems which can be identified with
the quantum isotonic oscillator \cite{Landau}. For certain values of parameters this can be
interpreted as a particle interacting through a spin-orbit interaction term with  an effective magnetic field.}
Such results are interesting also because
they show explicitly that the { ILC} actually determines
the form of the interaction term, and the dissipation controls the interaction strength.}

\section*{Acknowledgements}
%%%%%%%%%%%%%%%%%%%%%%%%%%%
%
The authors are grateful to G. 't Hooft for enlightening conversations.
F.S. thanks Japan Society for the Promotion of Science for support
under the fellowship P06782, National Taiwan University for support
under the research contract 097510/52313, and ITP Freie
Universit\"{a}t Berlin for warm hospitality. { P.J. would like to acknowledge
support by the Deutsche Forschungsgemeinschaft under
grant Kl256/47.} M.B. and G.V.
acknowledge partial financial support by INFN.
%
%
%
%
%%%%%%%%%%%%%%%%%%%%%%%%%%%%
\section*{Appendix 1}
%%%%%%%%%%%%%%%%%%%%%%%%%%%%
%
%
%
%********************
%
%In this
%connection it is perhaps useful to stress that in contrast to radial
%coordinates, in the hyperbolic coordinates $r_{A,B} \in
%{\mathbb{R}}$ and $p_{r_{A,B}} \in {\mathbb{R}}$ (see
%Eq.(\ref{hypm})).
%
%***********************

%

In this Appendix we comment more on the physical states
$|\psi\rangle_{phys}$ by closely following Ref.~\cite{celeghini92}.
We denote by ${\mathcal{H}}=\{| j_i,l_i\rangle, ~i = A,B\}$, the
states corresponding to simultaneous eigenstates of $\hat{J}_{3i}$
and $\hat{\cal C}_i$. These fulfill~\cite{Perelomov}:
\bea
&&\hat{J}_{3i} |j_i,l_i\rangle = \left(l_i + \mbox{$\frac{1}{2}$}\right)
|j_i,l_i\rangle\, , \label{III41a} \\[2mm]
&&\hat{\cal C}_i|j_i,l_i\rangle = j_i|j_i,l_i{\rangle\, .} \label{III41b}
\eea
The corresponding unitary irreducible representations (UIR's) can be
grouped into three independent non-overlapping classes (series)
according to the spectrum of $\hat{\cal C}_i$ and $\hat{J}_{3i}$.
Eigenvalues $l_i$ are in all three series real and discrete. On the
other hand, the operators $\hat{\cal C}_i$ and $\hat{J}_{2i}$
generate the so-called non-compact {\em hyperbolic} subgroup of
$su(1,1)$ (see, e.g., Refs.~\cite{barut77,lindblad70}).
%Because of the
%identity~\cite{celeghini92}
%%
%\bea \pm i \exp\left(\pm\frac{\pi}{2}\hat{J}_1\right) \hat{J}_3
%\exp\left(\mp \frac{\pi}{2}\hat{J}_1\right) \ = \ \hat{J}_2\, , \eea
%%
The corresponding UIR's fall into three series. However, the
spectrum of $\hat{J}_{2i}$ is either $\mathbb{R}\otimes
\mathbb{Z}_2$ or $\mathbb{R}$~\cite{lindblad70}. The latter depends
on the actual series.
%with $j_i = 0, \pm \frac{1}{2}, \pm 1, \pm \frac{3}{2}, \ldots$ and
%$l_i =|j_i|, |j_i| +1, |j_i| +2, \ldots $.
%In addition, according to
%our assumption only positive $j_i$ are considered.
%Here it is necessary to remark that one should be careful in handling the
%relation (\ref{III41b}) and states $|j_i,l_i\rangle$.
Besides UIR's, there exist also (non-unitary) representations in
which case  $\hat{J}_{2i}$ has discrete complex spectrum.
{ The usual
argument that a self-adjoint operator has only real eigenvalues does
not apply in this case because here}
one deals then with the extension of $\hat{J}_{2i}$ (that is
self-adjoint in a Hilbert space $\mathcal{H}$), to a larger space
$\mathcal{D} =\{| \Psi_{j_i,\mu_i} \rangle, ~i = A,B\}$,
\bea &&\hat{J}_{2i} |\Psi_{j_i,\mu_i} \rangle = \mu_i
|\Psi_{j_i,\mu_i} \rangle\, , \label{III41aa} \\[2mm]
&&\hat{\cal C}_i|\Psi_{j_i,\mu_i} \rangle = j_i|\Psi_{j_i,\mu_i}
\rangle\, , \label{III41c} \eea
in which $\mathcal{H}$ is dense. It can be shown~\cite{lindblad70}
that such an extension is nothing but the closure by continuity of
continuous operators defined originally on the dense subspace
${\mathcal{H}}$.
%It is sometimes useful to rectify this ``pathology"
%by endowing the space ${\mathcal{D}}$ with a suitable inner
%product~\cite{celeghini92,blasone3} and hence promoting it formally
%to a new Hilbert space. From the mathematical point of view the
%natural language to analyze the spectral properties of Bateman's
%system is the rigged Hilbert space or Gel'fand
%triplet~\cite{lindblad70,chruscinski}. We shall, however, not dwell
%on this point further.

The choice of a representation { corresponding to physical Hilbert
space $\mathcal{O}$ is tightly} connected with the fact
that the state space of the underlying be-able system is
${\mathcal{D}}\otimes{\mathcal{D}}$ and that the spectrum of
$\hat{H}_i$ should correspond to poles of the resolvent operator
$R_z(\hat{H}_i) = (\hat{H}_i - z)^{-1}$. This restricts to a
non-unitary realization of the discrete principal series
$D_j$~\cite{celeghini92,chruscinski,fesch-tich}. The series $D_j$
comes in two copies,
%~\footnote{Two copies of $D_j$
%result from an outer automorphism of the $su(1,1)$
%algebra~\cite{lindblad70}: $(J_1,J_2,J_3) \mapsto (-J_1, J_2,
%-J_3)$. This automorphism can be realized with an operator
%$\hat{P}$, $\hat{P}^2 =1$. In particular, $[\hat{P}, \hat{J}_2] =0$
%which implies that both $\hat{P}$ and $\hat{J}_2$ can be
%diagonalized simultaneously.}
namely $D_j^{+}$ and $D_j^{-}$ (where $D_j^{-}\cap D_j^{+} =
\{\not\!\!0 \}$). It was shown in
Refs.~\cite{celeghini92,blasone3,chruscinskiII} that for BO's
$D_j^{+}$ corresponds to the forward-time dynamics while $D_j^{-}$
describes the backward-time dynamics. Denote the {\em generalized}
eigenvectors belonging to $D_j^{\pm}$ as $|\Psi^{\pm}_{j,\mu}
\rangle$, then~\cite{lindblad70,celeghini92}
\bea &&\hat{J}_{2i} |\Psi^{\pm}_{j_i,\mu_i}\rangle = \pm\mu_i
|\Psi^{\pm}_{j_i,\mu_i}\rangle\, , \label{III42aa} \\[2mm]
&&\hat{\cal C}_i|\Psi^{\pm}_{j_i,\mu_i}\rangle =
j_i|\Psi^{\pm}_{j_i,\mu_i}\rangle\, , \label{III42c} \eea
with $j_i = 0, \pm \frac{1}{2}, \pm 1, \pm \frac{3}{2}, \ldots$ and
$\mu_i = i(l_i + \frac{1}{2})$,  $l_i =|j_i| + m_i$, $m_i\in
{\mathbb{N}}$. There exits a natural outer automorphism between
$D_j^{+}$ and $D_j^{-}$ which can be on a physical level identified
with the time reversal
operation~\cite{celeghini92,blasone3,chruscinski}, in particular
\begin{eqnarray}
D_j^{+} = \mathcal{T}(D_j^{-})\, , \;\;\;\;\; D_j^{-} =
\mathcal{T}(D_j^{+})\, ,
\end{eqnarray}
where $\mathcal{T}$ is the time reversal operator.
%From a
%mathematical point of view correspond the generalized eigenvectors
%$|\Psi^{\pm}_{j_i,\mu_i} \rangle $ to resonant states of the
%Bateman's dual system.

By imposing the constraint (\ref{III.39}) one can identify the
physical states $|\psi\rangle_{phys}$ with states
\begin{eqnarray}
 \left\{|\psi^{+}_{j_A,\mu_A}\rangle \oplus
|\psi^{-}_{j_A\mu_A}\rangle \right\} \otimes
\left\{|\psi^{+}_{j_B,\mu_B}\rangle \oplus
|\psi^{-}_{j_B,\mu_B}\rangle \right\}\, .
\end{eqnarray}
%
%correspond to a non-unitary realization of the $D^+_{1/2}$
%representation of $su(1,1)$.
Here $|\psi^{\pm}_{j_i,\mu_i}\rangle$ belongs to a two-dimensional
space spanned by vectors $\{|\Psi^{\pm}_{-j_i,\mu_i}\rangle,
|\Psi^{\pm}_{j_i,\mu_i}\rangle \}$.
%
%
%
%
%
%%%%%%%%%%%%%%%%%%%%%%%%%%%%
\section*{Appendix 2}
%%%%%%%%%%%%%%%%%%%%%%%%%%%%
%
%
%
%
By considering $m_A = m_B = M$, $\Omega_A = \Omega_B = \Omega$,
(this can be achieved by properly rescaling canonical variables and
Hamiltonian), we rewrite (\ref{66}) as
\begin{eqnarray}
\H_T  \ \approx \ \H_+ &\approx& \frac{{\hat p}_{r_A}^2 +
{\hat p}_{r_B}^2}{2M}+\frac{1}{2}M\Omega^2(\hat{r}_A^2 +
\hat{r}_B^2)\nonumber \\[2mm]
 &&- \frac{2}{M}\left(\frac{1}{\hat{r}_A^2} +
\frac{\Gamma_A^2}{\Gamma_B^2}\frac{1}{\hat{r}_B^2} \right)
\hat{J}_{2A}^2 \, , \label{IV56}
\end{eqnarray}
which is known as the two-dimensional Smorodinsky--Winternitz
system~\cite{Grosche}. It belongs to a class of two-dimensional
maximally super-integrable models~\cite{fris65}. Through its
centripetal term, the system (\ref{IV56}) is related to the
Calogero--Moser system~\cite{Calogero}.

The spectrum of (\ref{IV56}) can be written in the form~\cite{Grosche}\
\begin{eqnarray}
E_{n_A,n_B}(\mu_A) \ &=&\ \Omega \left(2(n_A + n_B) \ \pm \
\sqrt{\frac{1}{4} - 4\mu^2_A}\right.
\nonumber \\[2mm] &&\pm \left. \sqrt{\frac{1}{4} - \left(\frac{2\mu_A
\Gamma_A}{\Gamma_B}\right)^{\!2}} + \ 2\right)\, , \label{66b}
\end{eqnarray}
where $n_A,n_B \in {\mathbb{N}}$. Note that for the attractive
potential both signs must be taken into account, while for the
repulsive one only ``$+$" sign counts.

When $\mu_A$ is a real number in (\ref{66b}), i.e., when
$\hat{J}_{2A}$ belongs to a unitary realization of the $D_j$ series,
then both potentials in (\ref{IV56}) are  attractive. As a result
the motion takes place in the domain $(r_B,r_A) \in \mathbb{R}^2$
and both signs in (\ref{66b}) must be taken into account. To avoid
``fall'' to the center~\cite{Landau} both square roots in
(\ref{66b}) must be real numbers. This implies that $\mu_A\in
[-1/4,1/4]$ and restricts the possible values of $\Gamma_A/\Gamma_B$
to the interval $[-1,1]$.


\begin{thebibliography}{99}

\bibitem{hooft1}
G.~'t~Hooft, J. Stat. Phys. {\bf 53} (1988) 323; G.~'t~Hooft, Class.
Quant. Grav. {\bf 16}, 3263 (1999) [gr-qc/9903084]; G.~'t~Hooft,
{\em Determinism beneath Quantum Mechanics} [quant-ph/0212095];
G.~'t~Hooft, Class. Quant. Grav. {\bf 13} (1996) 1023.
%
\bibitem{hooft2}
G.~'t~Hooft, in 37th International School of
Subnuclear Physics, Erice, Ed. A. Zichichi, (World Scientific,
London, 1999) [hep-th/0003005];
G.~'t~Hooft, Int. J. Theor. Phys. {\bf 42} (2003) 355
[hep-th/0105105] .
%
%
\bibitem{Biro}
T.~S.~Biro, S.~G.~Matinyan and B.~Muller,
Found. Phys. Lett. {\bf 14} (2001) 471. [hep-th/0105279]
%
%
%\bibitem{vandeBruck:2000cw}
%C.~van de Bruck,
%``On gravity, holography and the quantum,''
%[gr-qc/0001048].
%%CITATION = GR-QC 0001048;%%
%
\bibitem{benerjee}  R.~Banerjee, Mod. Phys. Lett. A {\bf 17} (2002) 631
[arXiv:hep-th/0106280].
%
\bibitem{elze} H.-T.~Elze,  J. Phys.: Conf. Ser.
33 (2006) 399.
%
\bibitem{elze2} H.-T.~Elze,
Int. J. Quantum Info. (IJQI) {\bf 7} (2009) 83 [arXiv:0806.3408].
%
\bibitem{halliwell} J.J.~Halliwell, Phys. Rev. D {\bf 63} (2001)
085013.
%
%
\bibitem{blasone1}
M.~Blasone, P.~Jizba, G.~Vitiello, Phys. Lett. A {\bf 287}, 205
(2001).
%
%
\bibitem{blasone2}
M.~Blasone, P.~Jizba and G.~Vitiello, J. Phys. Soc. Jap. Suppl. {\bf
72} (2003) 50; M.~Blasone and P.~Jizba, Can. J. Phys. {\bf 80}
(2002) 645; M.~Blasone, E.~Celeghini, P.~Jizba and G.~Vitiello,
Phys. Lett. A {\bf 310} (2003) 393.
%
%
\bibitem{scard}
F.~Scardigli, Found. of Phys., {\bf 37}, 1278 (2007).
%
%
\bibitem{Jizba}
M.~Blasone, P.~Jizba and H.~Kleinert, Phys. Rev. A {\bf 71} (2005)
052507.
%
\bibitem{JizbaII} M.~Blasone, P.~Jizba and H.~Kleinert,
Ann. Phys. {\bf 320} (2005) 468.
%
\bibitem{JizbaIII} M.~Blasone and P.~Jizba, J. Phys.: Conf. Ser. {\bf 67} (2007)
012046.
%
%

%
\bibitem{celeghini92} E.~Celeghini, M.~Rasetti and G.~Vitiello, Ann. Phys. {\bf 215} (1992) 156.
%

\bibitem{Landau} L.D.~Landau and E.M.~Lifshitz, {\em Quantum Mechanics: Non-relativistic Theory}
(Pergamon, New York, 1977).

%\bibitem{Umezawa:1985} H Umezawa and G. Vitiello, {\em Quantum
%Mechanics} (Bibliopolis, Napoli 1985).
%

%
%
\bibitem{pontryagin62} L.S.~Pontryagin, V.G.~Boltanskij,
R.V.~Gamkrelidze and E.F.~Miscenko, {\em The Mathematical Theory of
Optimal Processes} (Wiley, New York, 1962).
%
%

\bibitem{foot1} Hermiticity of $\H$ is not
a problem at the energy scale $E_P$, since the Hamiltonian is only a formal tool
useful to generate evolution equations. However, one can always compensate for
non-hermitian ordering by adding the non dynamical arbitrary function $g({\hat{\bf q}})$
in such a  way that $g^\dagger ({\hat{\bf q}})-g({\hat{\bf q}}) = \sum_i [\hat{p}_i , f_i({\hat{\bf q}})]$
so that $\H = \H^\dagger$.

\bibitem{Bell} J.S.~Bell, {\em Speakable and Unspeakable in Quantum
Mechanics} (Cambridge University Press, Cambridge, 1987).

\bibitem{sunder} see, e.g.,
K.~Sundermeyer, {\em Constrained Dynamics with Applications to
Yang-Mills theory, General Relativity, Classical Spin, Dual String
Model} (Springer Verlag, Berlin, 1982); M.~Henneaux and
C.~Teitelboim, {\em Quantization of Gauge Systems} (Princeton
University Press, Princeton, 1991); P.A.M.~Dirac, Can. J.Math.  2 (1950)
129; Proc. R. Soc. London.  A {\bf 246} (1958) 326; {\em Lectures
in Quantum Mechanics} (Yeshiva University, New York, 1965).
%
\bibitem{BlasVit} M.Blasone, E.Graziano, O.K.Pashaev, G.Vitiello, Ann. Phys. {\bf 252} (1996) 115.




\bibitem{Perelomov} A.~Perelomov, {\em Generalized Coherent States and Their Applications}
(Springer-Verlag, London, 1986).
%
\bibitem{Wang00} J.S.~Wang, T.K.~Liu and M.S.~Zhan, J. Optics B {\bf 2} (2000)
758.
%
\bibitem{Saad04} K.~Thirulogasanthar and N.~Saad, J. Phys. A {\bf 37} (2004)
4567.

\bibitem{Dongpei} Z.~Dongpei, J. Phys. A.: Math. Gen. {\bf 20} (1987)
4331.
%
\bibitem{Grosche} C.~Grosche, G.S.~Pogosyan and A.N.~Sissakian,
Fortschr. Phys. {\bf 43} (1995) 6.
%

\bibitem{Calogero} see, e.g., E.~Corrigan and R.~Sasaki,
J.Phys. A {\bf 35} (2002) 7017  and citations therein.
%

\bibitem{MS} J.~Magueijo, L.~Smolin, Phys. Rev. D {\bf 67} (2003) 044017.


\bibitem{foot2} This form is
inspired by renormalization group considerations and the information entropy
form, and we plan to further explore it in future work.

\bibitem{barut77} A.~Barut and R.~Raczka, {\em Theory of Group Representations and
Applications} (Polish Scientific Publisher, Warszawa, 1977).
%
\bibitem{lindblad70} G.~Lindblad and B.~Nagel, Ann. Inst. Henri
Poincar\'{e} Sect. A {\bf 13} (1970) 27.
%
%
%
\bibitem{fesch-tich} H.~Feshbach and Y.~Tikochinski, Trans. N.Y. Acad. Sci., Ser. II
{\bf 38} (1977) 44.
%

%
\bibitem{chruscinski} D.~Chruscinski, Ann. Phys. {\bf 321} (2006) 840.

\bibitem{blasone3} M.~Blasone and P.~Jizba, Ann. Phys. {\bf 312} (2004) 354.
%
\bibitem{chruscinskiII} D.~Chruscinski, J.~Jurkowski, Ann. Phys. {\bf 321} (2006) 856.
%
%
%

%
\bibitem{fris65} J.~Fri\u{s}, V.~Mandrosov, Ya.A.~Smorodinsky,
M.~Uhl\'{i}\u{r} and P.~Winternitz, Phys. Lett. {\bf 16} (1965) 354.
%
%%
%
%
% \bibitem{Lichnerowicz}
%  F.~Bayen,  M.~Flato, C.~Fronsdal, A.~Lichnerowicz, and D.~Sternheimer, Ann.~Phys {\bf 111} (1978) 61;
%  ibid.{\bf 111} (1978) 111.
% %
%

%


\end{thebibliography}
\end{document}